\documentclass[preprint]{ptephy_v1}

\preprintnumber{} 

\usepackage{amsmath} 
\usepackage{amsthm} 
\usepackage{graphics} 
\usepackage{algorithmic} 

 \def\m@thcombine#1#2{%
  \setbox0=\hbox{$#1$}
  \setbox1=\hbox{$#2$}
  \ifdim\wd0>\wd1
    \setbox0=\hbox to\wd1{\hss\box0\hss}
  \else
    \setbox1=\hbox to\wd0{\hss\box1\hss}
  \fi
  \mathop{\vcenter{
    \offinterlineskip\box0\box1}}}
\def\lesim{\m@thcombine<\sim}
\def\gesim{\m@thcombine>\sim}

\begin{document}

\title{Proximity effect of pair correlation in the inner crust of neutron stars}

\author[1]{Toshiyuki Okihashi}
\affil{Graduate School of Science and Technology, Niigata University, Niigata 950-2181, Japan}

\author[2,*]{Masayuki Matsuo}
\affil{Department of Physics, Niigata University, Niigata 950-2181, Japan \email{matsuo@phys.sc.niigata-u.ac.jp}}

\begin{abstract}%
We study proximity effect of pair correlation in the inner crust of neutron stars by
means of the Skyrme-Hartree-Fock-Bogoliubov theory formulated in the coordinate space.
We describe a system composed of a nuclear cluster immersed in neutron superfluid,
which is confined in a spherical box. 
Using 
a density-dependent effective pairing interaction 
which reproduces  both the pair gap of  neutron matter
obtained in ab initio calculations and that of finite nuclei,
we analyze how the pair condensate in neutron
superfluid  is affected by the presence of the nuclear cluster.
It is found
that the proximity effect is characterized by the
coherence length of neutron superfluid measured from the edge position of the nuclear 
cluster. The calculation predicts that the proximity effect has a strong density dependence.
 In the middle layers of the
inner crust with baryon density $5 \times 10^{-4}$ fm$^{-3} \lesim \rho_b \lesim 2\times 10^{-2}$ fm$^{-3}$,
the proximity effect
is well limited in the vicinity of the nuclear cluster, i.e. in a sufficiently smaller area than
the Wigner-Seitz cell. 
On the contrary, the proximity effect is predicted to
extend to the whole volume of the Wigner-Seitz cell in shallow layers of
the inner crust with $\rho_b \lesim 2 \times 10^{-4}$ fm$^{-3}$, 
and in deep layers with $\rho_b \gesim 5 \times 10^{-2}$ fm$^{-3}$.
\end{abstract}

\subjectindex{D10,D11,D13,D41}

\maketitle

\section{Introduction}
The inner crust of neutron stars is an exotic inhomogeneous matter consisting of a lattice of
neutron-rich nuclear clusters which is immersed in neutron superfluid \cite{Chamel-Haensel2008}. One of the
central issues of the physics of the inner crust is interplay between
the superfluidity and the inhomogeneity, which  influences
various properties of the inner crust such as
the specific heat, the thermal conductivity, and the pinning and unpinning of 
vortices. These are essential factors to understand astrophysical issues, such as the cooling
and the glitch phenomenon of the neutron stars. 

Microscopic many-body approaches to these phenomena have been
pursued in the framework of  the  Hartree-Fock-Bogoliubov (HFB) theory, which has a capability
to describe microscopically the inhomogeneous pair-correlated system. 
It has been argued for instance that
the presence of the nuclear cluster modifies
 the quasiparticle excitation spectrum and the
average pair gap, leading to a sizable difference in the specific heat of the
inner crust from that of the uniform neutron superfluid
\cite{Pizzochero2002,sandulescu2004,Monrozeau2007,chamel2010,Pastore2015}. 
evaluate the pinning energy of superfluid vortices \cite{Avogadro2008,Wlazlowski2016,Jin2017}. 
Recent interest also
concerns with a dynamical aspect of the issues, i.e. 
the interaction between the vibrational motion of the nuclear cluster and
the phonon excitation (the Anderson-Bogoliubov collective mode) of the neutron superfluid.
This is one of the key ingredients which influence the thermal conductivity of the inner crust 
in magnetars \cite{aguilera2009,Pethick10,Cirigliano11,Page-Reddy2012,Chamel13}. 
In an attempt to analyze this dynamical coupling from a microscopic viewpoint, we
have investigated the collective excitation of the inner crust matter by means
of  the quasiparticle random phase approximation based on the 
HFB theory \cite{inakura2017,inakura2019}. We found that the dynamical coupling
between the collective motions of the nuclear cluster  and of the neutron
superfluid is weak.
  
In the present study, we intend to reveal  the interplay between  the
nuclear cluster and the neutron superfluid but from a different viewpoint,
i.e. the proximity effect of the pairing correlation \cite{gennes1964,gennesbook}.
The proximity effect is a general phenomenon which emerges around a
border region of the system of a superconducting/superfluid
matter in contact with normal matter (or matter with different pairing property). 
The pairing correlations in both matter are affected mutually in the border
region since the Cooper pairs penetrate the border. 
The proximity effect in the inner crust matter is discussed in a few preceding
works \cite{chamel2010,sandulescu2004,barranco1998} , but only in a qualitative manner. In the present study, we aim at 
characterizing the proximity effect quantitatively in order to reveal
basic features of the pair correlation arising from the inhomogeneous
structure of the inner crust matter.   

As a theoretical framework to perform this study, 
we adopt the HFB theory using the Skyrme functional with a implementation
of a few new features. 
One of the key elements in the HFB approach is the effective pairing interaction or
the effective pairing functional, which generates the pair correlation in the system
under study, and a density-dependent contact force, called the density-dependent delta
interaction (DDDI), is often adopted.  Note however that
the inner crust matter consists  of the neutron superfluid, whose density
varies in a wide range from zero to that of the nuclear saturation, and 
the nuclear clusters, which resemble to isolated neutron-rich nuclei. 
In order to take into account this feature, 
we prepare a new parameter set of DDDI, which is required to describe the
pairing gap of neutron superfluid obtained in ab initio calculations \cite{gezerlis2010,abe2009} 
as well as the experimental pairing gap in finite nuclei.
Secondly, we quantify the range of the proximity effect by identifying  the distance where
the  presence of the nuclear cluster  influences the pairing property in neutron
superfluid. Using this measure, we discuss in detail the dependence of the
proximity effect on the density of the neutron superfluid, and clarify how
large the proximity effect is in different layers of the inner crust.

In Section 2, we explain the adopted Skyrme-HFB model and the new parameter set 
of DDDI. In the present HFB all the nucleons are described as quasiparticles confined in a 
spherical box. If we adopt the box size equal to the Wigner-Seitz radius of the
lattice cell, it is the same as the Wigner-Seitz approximation often adopted in
the preceding works. However, the box truncation causes so called  finite-size
effect, and it  make difficult to analyze the proximity effect.
In Section 3, we examine  the finite-size effect, and propose a different
setting of the analysis using a large box truncation in place of the Wigner-Seitz approximation.
Section 4 is devoted to a systematic analysis of the proximity effect. In subsection
4.1 we describe our scheme of the analysis that quantifies the range of the proximity effect, and justify
the scheme with a systematic variation of the density of neutron superfluid immersing the nuclear
cluster. In subsection 4.2, we apply the same analysis to various layers of a
realistic configuration of the inner crust of neutron stars.
Section 5 is devoted to the conclusions.

 \section{Model}
 
 \subsection{Skyrme-Hartree-Fock-Bogolibov method in a spherical box} 
 
We adopt the Skyrme-Hartree-Fock-Bogoliubov method to describe the inner crust matter.
Since the method is an extension of that is used in Refs. \cite{inakura2017,inakura2019}, we describe it
 briefly with emphasis on new aspects which are introduced in the present study.

We solve the HFB equation in a spherical box using the radial coordinate space 
and the partial wave expansion. The zero temperature is assumed and the spherical symmetry 
of solutions is imposed. Electrons are neglected. 
The radial HFB equation for a given angular quantum numbers $lj$ reads
\begin{align}
\begin{bmatrix}
 h_{qlj}(r)-\lambda_q&\Delta_q(r)\\
 \Delta_q(r)&-h_{qlj}(r)+\lambda_q
\end{bmatrix}
 \begin{bmatrix}
  \phi_{1}^{qlj}(r)\\
    \phi_{2}^{qlj}(r)
 \end{bmatrix}
 =E
 \begin{bmatrix}
  \phi_1^{qlj}(r)\\
  \phi_2^{qlj}(r)
 \end{bmatrix}
  ,
   \label{HFBeq}
 \end{align}
 where $\phi_1^{qlj}, \phi_2^{qlj}$ is the quasiparticle wave function. 
 Index $q$ denotes neutron or proton.

 We discretize the radial coordinate with an
interval $h=0.2$\,fm  as $r_i=i*h-h/2=h/2, 3h/2,  \cdots$ $(i=1,\cdots,N)$ up to 
the edge $r=R_{\mathrm{box}}$ of the box, and use  the nine-point formula 
to represent the derivatives in the Hartree-Fock Hamiltonian $h_{qlj}(r)$.
 We impose the Dirichlet-Neumann boundary condition \cite{NV1973}, with which
 even-parity wave functions vanish at the edge of the box and
 the first derivatives of odd-parity wave functions vanish at the same position. 
  Equation (\ref{HFBeq}) is represented as a matrix eigenvalue problem where
 the wave function at the discretized coordinates 
 $\left( \phi_{1}^{qlj}(r_1),\cdots \phi_{1}^{qlj}(r_N), \phi_{2}^{qlj}(r_1),\cdots \phi_{2}^{qlj}(r_N)\right)^{T}$ is 
  a $2N$-dimensional vector.
 We use routine DSYEVX in the LAPACK package to solve the eigenvalue problem
 for the symmetric matrix.
 If we treat the
 lattice configuration of the nuclear clusters by means of the Wigner-Seitz approximation, 
 the box radius $R_\mathrm{box}$ is chosen to be the size of the Wigner-Seitz cell. 
 We shall also choose larger boxes   $R_\mathrm{box}=100$ fm
 or  200 fm, as we explain below. 
  All the quasiparticle states up to a maximal quasiparticle energy $E_\mathrm{max}=60$ MeV  are
 included to calculate the number density, the pair density and all the quantities needed to calculate the selfconsistent
 potentials.   We put also a cut-off $l_\mathrm{max}$ on the angular momenta of the partial waves so
 that  $l_\mathrm{max}>\sqrt{E_{\mathrm{max}}/(\hbar^2/2m)}R_\mathrm{box}$: 
 $l_\mathrm{max}=200\hbar$ for $R_\mathrm{box}=100$ fm,  and 
 $l_\mathrm{max}=400\hbar$ for $R_\mathrm{box}=200$ fm, for example.
We use the parameter set SLy4 \cite{chabanat} for the selfconsistent Hartree-Fock potential in $h_q(r)$.  We adopt 
the density-dependent delta interaction, as described below, to derive the pair
potential  $\Delta_q(r)$. 
 We vary the neutron Fermi energy $\lambda_n$ to control the neutron density
 and we determine the proton Fermi energy $\lambda_p$ to fix the proton number $Z$ of the nuclear
 cluster. 
 The other details are the same as in the previous study \cite{inakura2017,inakura2019}.

 \subsection{Density-dependent pairing interaction}

 As the pairing interaction, we use a density-dependent delta-interaction (DDDI), given as
 \begin{align}
 v_{\mathrm{pair},n}(\vec{r}_1,\vec{r}_2)=V_n[\rho_n(\vec{r}),\rho_p(\vec{r})]
 \left(\frac{1-P_\sigma}{2}\right)\delta(\vec{r}_1-\vec{r}_2), 
 \hspace{5mm} \vec{r}=\vec{r}_1 (=\vec{r}_2),
 \label{DDDI}
\end{align}
 for neutrons. Here $V_n[\rho_n(\vec{r}),\rho_p(\vec{r})]$ is the density-dependent interaction strength, 
 and  $(1-P_\sigma)/2$ is the projection operator for the spin singlet channel.
 The pair potential is then $\Delta_n(r)= V_n[\rho_n(r),\rho_p(r)]\tilde{\rho}_n(r)$  with
 the neutron pair density (the neutron pair condensate)
 \begin{align}
 \tilde{\rho}_n(\vec{r})= \langle \psi_n(\vec{r}\uparrow)\psi_n(\vec{r}\downarrow)\rangle.  
 \end{align}
 We consider
 the following three models for the interaction strength $V_n[\rho_n(\vec{r}),\rho_p(\vec{r})]$.
 
 The first one, which we introduced in Refs.~\cite{matsuo2006,matsuo2007}, is given as
 \begin{equation}
 V_n[\rho_n(\vec{r})]= 
 V_0\left\{1-0.845\left(\frac{\rho_n(r)}{\rho_0}\right)^{0.59}
 \right\}
 \label{DDDI-strong}
\end{equation}
with $\rho_0=0.08$ fm$^{-3}$. Here the overall constant 
$V_0=-458.4$ $\mathrm{MeV\,fm^{3}}$ is determined to
 reproduce the ${}^1S_0$ scattering length $a$=-18.5 fm in free space (i.e. at 
 zero density) under the single-particle cut-off energy $e_\mathrm{cut}=60$ MeV.
  The dependence of the interaction strength $V_n[\rho_n]$ on the neutron density $\rho_n$ is determined
  so that it reproduces the neutron pairing gap in pure neutron matter which is obtained in the BCS
 approximation using a bare nuclear force \cite{matsuo2006,matsuo2007}.
  We denote the parameterization, Eq. (\ref{DDDI-strong}),
 as ``DDDI-b'' since it refers to the {\it B}CS gap with the {\it b}are nuclear force. 
 (It is the same as the parametrization DDDI-G3RS in Ref. \cite{matsuo2006}.)

 \begin{table}[t]
  \centering
    \caption{DDDI parameters adopted in the present study. For the definition, see Eq. (\ref{DDDI-med})
    and the text. The parameters are appropriate for the cut-off energy $e_\mathrm{cut}=60$ MeV.}
  \begin{tabular}{lcccccccc}
   \hline
   &$V_0\,(\mathrm{MeV fm}^{3})$&$\rho_0\,(\mathrm{fm^{-3}})$&$\eta_1$&$\alpha_1$&$\eta_2$&$\alpha_2$&$\eta'_1$&$\alpha'_1$\\\hline
   DDDI-a1&-458.4&0.08&0.59&1/3&0.06&2/3& 0 & 1/3\\
   DDDI-a2&-458.4&0.08&0.59&1/3&0.255&2/3 & -0.195 & 1/3\\
      DDDI-b&-458.4&0.08&0.845&0.59& - & - & - & -\\
\hline
  \end{tabular}
  \label{t1}
 \end{table}

 In the present study we introduce more realistic modeling of the neutron pairing appropriate to the inner crust matter. Here we consider parametrizations of the DDDI that 
 provide realistic pairing gap both in neutron matter and in finite nuclei.
 Concerning the neutron matter, it is known that the pairing gap is affected by medium effects
 beyond the BCS approximation,
 and many of theoretical studies trying to evaluate the
 medium effects predict a significant reduction from the
 BCS gap while the predicted values spread in a wide range \cite{Dean2003,Lombardo2001,Gandolfi-Gezerlis-Carlson,Strinati2018}.
Nevertheless, the pairing gap in the low-density limit
is believed to be described reliably by a perturbative approach to the screening effect,  
discussed first by Gor'kov and Melik-Barkhudarov (GMB) \cite{gmb}, and the pairing gap 
$\Delta_\mathrm{GMB}$ in the GMB framework gives a reduction of a factor of $(4e)^{1/3}\simeq 2.2$ from the BCS
pairing gap\cite{heiselberg2000,Pethick-Smith,Strinati2018}. Recently, numerical ab initio calculations
based on Monte-Carlo methods have been performed for pure neutron matter
in low density region $\rho_n \simeq 10^{-5} - 10^{-2}$ fm$^{-3}$,
and the predicted
pairing gaps are reduced from the BCS gap by a factor of 1.5 - 2 \cite{gezerlis2010,abe2009,Gandolfi2008,Gandolfi-Gezerlis-Carlson}. We can refer to these
studies in requiring a new parametrization of the DDDI.  It is also known that the pairing gap in
finite nuclei cannot be described well by the BCS approximation applied to the bare nuclear force,
and there is no ab initio evaluation of the gap in finite nuclei.
Instead we will refer to experimental information on the pairing gap in finite nuclei.

In order to satisfy these conditions we introduce the following extended form of the density-dependent 
interaction strength:
\begin{equation}
V_n[\rho_n(\vec{r}),\rho_p(\vec{r})]=
V_0\left\{1-\eta_1\left(\frac{\rho_n(r)}{\rho_0}\right)^{\alpha_1}
 -\eta_2\left(\frac{\rho_n(r)}{\rho_0}\right)^{\alpha_2}
-\eta'_1\left(\frac{\rho_p(r)}{\rho_0}\right)^{\alpha'_1}
 \right\}.
 \label{DDDI-med}
 \end{equation}
 
 The first term is introduced to describe the GMB gap appropriate to the low-density
limit of the pure neutron matter.
As discussed in Appendix A, the force strength $V_\mathrm{GMB}$ of the contact force
which reproduces the GMB pairing gap $\Delta_\mathrm{GMB}$ depends on the neutron
Fermi momentum $k_{F,n}$ or the density $\rho_n$ of neutron matter. The dependence is
expressed as  a
linear term proportional to $k_{F,n}$ or $\rho_n^{1/3}$ if it is expanded in powers of $k_F$.
Requiring that the GMB pairing gap is reproduced by the DDDI 
in the low-density limit $\rho_n \rightarrow 0, k_{F,n} \rightarrow 0$.
the parameters of the first term in Eq. (\ref{DDDI-med}) is fixed to $\alpha_1=1/3$ and $\eta_1=0.59$. 

The second and third terms are introduced to represent the pairing gap of neutron matter
at finite density and that in finite nuclei.  In particular,
the second term together with the first term is relevant to the pairing gap in neutron matter,
and we assume that the second term has 
a power $\alpha_2=2/3$, i.e., $\propto \rho_n^{2/3}\propto k_{F,n}^2$ the second power of neutron Fermi momentum $k_{F,n}$. 
We then require that
the coefficient $\eta_2$ of this term is  consistent with
the ab initio pairing gap of neutron matter
obtained  for $10^{-5}$ fm$^{-3} \lesssim \rho_n \lesssim 10^{-2}$ fm$^{-3}$ in
the quantum Monte Carlo calculation by Gezerlis and Carlson \cite{gezerlis2010} and the
determinantal lattice Monte Carlo calculation by Abe and Seki \cite{abe2009}. 
Note however that this requitement alone does not fix uniquely the coefficient $\eta_2$ 
since these ab initio calculations are slightly different with each other and
there is no ab initio results for moderately low densities
$10^{-2}$~fm$^{-3} \lesssim \rho_n \lesssim 10^{-1}$~fm$^{-3}$. 

The third term dependent on the proton density 
represents a part of medium effects associated with systems with a proton fraction.
For simplicity we assume that it is
proportional to the proton Fermi momentum $k_{F,p}$ or the proton density $\rho_p^{1/3}$.
\footnote{
A perturbative estimate of the medium effect in symmetric matter gives an attractive induced interaction
proportional to $N_{0,p} \propto k_{F,p}$ \cite{heiselberg2000}. }  
We use both the coefficient $\eta'_1$ of this term and the uncertainty in  $\eta_2$
to describe the pairing gap in finite nuclei. 
 In practice, we require that  the average neutron pairing gap  $\Delta_{n,uv}=\int \Delta_n(r)\rho(r)d\vec{r}/\int\rho(r)d\vec{r}$
 in $^{120}$Sn 
 obtained from  our HFB model reproduces
 the experimental neutron
gap $\Delta_{n,\mathrm{exp}}\simeq 1.3 $ MeV, extracted from the 3-point odd-even mass difference \cite{Satula1998}.

\begin{figure}[t]
 \centering
 \includegraphics[width=0.6\hsize, bb = 0 0 300 200]{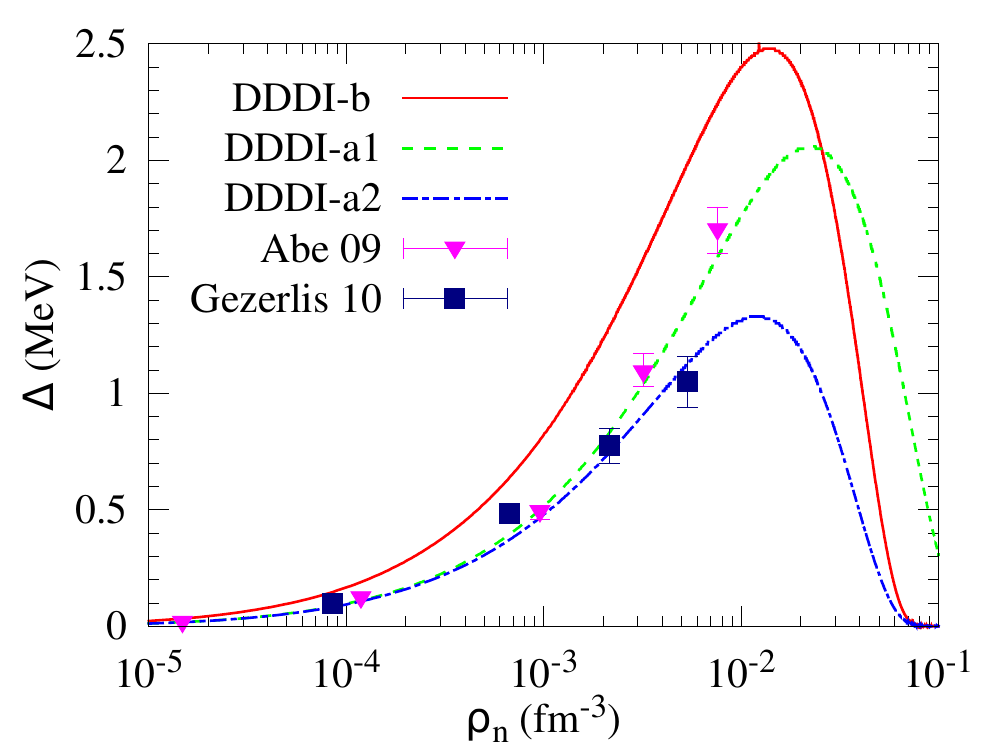}
    \caption{The pair gap $\Delta$ of superfluid neutron matter obtained in the uniform-BCS calculation 
    using the three DDDI models, plotted as a function of neutron density. 
    Solid, dashed and dot-dashed curves represent results for DDDI-b, DDDI-a1 and DDDI-a2, respectively.
    The ab initio Monte-Carlo results by  Gezerlis and Carlson \cite{gezerlis2010}
    and Abe and Seki \cite{abe2009} are shown for comparison.}
   \label{fig1}    
\end{figure}

In the present study 
we prepare two different parameter sets to represent the remaining uncertainty of the neutron pair gap.
In one case (we call ``DDDI-a1" below), we choose $\eta_2=0.06$ and $\eta'_1=0$ so that the
neutron pairing gap in $^{120}$Sn is reproduced without $\eta'_1$. In this case, 
the pairing gap of neutron matter is close to that of
Abe and Seki \cite{abe2009}, and the neutron matter pairing gap at moderately low density is rather large 
$\Delta \sim 1-2$ MeV, as shown in Fig.~\ref{fig1}.
It is remarked that the medium effect associated with the nuclear cluster or finite nuclei 
is effectively included in $\eta_2$. In another parameter set (``DDDI-a2"), we consider a case
that the neutron matter pairing gap at moderately low density is relatively small; we determine $\eta_2=0.255$ so
as to make   the neutron matter pairing gap vanish at
$\rho_n =\rho_0$ as the BCS gap does. The parameter $\eta'_1=-0.195$ is then determined to
reproduce the neutron gap in $^{120}$Sn. (Note that the neutron matter pairing gap 
reproduces approximately  the result of
Gezerlis and Carlson \cite{gezerlis2010}, as shown in Fig.~\ref{fig3}.)
The parameter sets of the three DDDI models are summarized in Table~\ref{t1}.

Figure \ref{fig1} shows the neutron pairing gap in uniform neutron matter obtained from the Hartree-Fock
plus BCS calculation using the DDDI models discussed above and the Skyrme functional
with the parameter set SLy4. (The BCS calculation is briefly recapitulated in Appendix B,
and we call it the uniform-BCS calculation in the following.)
By construction, the pairing gaps obtained with the DDDI-a1 and the DDDI-a2 reproduce
reasonably well the gap obtained with the ab initio calculations. The DDDI-a2
reproduces approximately  the result of
Gezerlis and Carlson \cite{gezerlis2010} for the density range $\rho_n=10^{-5}-10^{-2}$ fm$^{-3}$. The gap of the DDDI-a2 at 
moderate density is small $ \Delta < 1.3$ MeV, 
and vanishes at  $\rho_n \sim 0.08$ fm ($k_F \sim 1.4$ fm$^{-1}$)  corresponding
to neutrons in the saturated nuclear matter. The neutron gap of the DDDI-a1 is  very close to that of
the DDDI-a2 up to $\rho_n \lesim 10^{-3}$ fm$^{-3}$, 
but deviate from it above $\rho_n \gesim 10^{-3}$ fm$^{-3}$, It is rather close to the gap of 
Abe and Seki \cite{abe2009}, and the neutron matter pairing gap at moderately low density is rather large 
$\Delta \sim 1-2$ MeV.
The parameter set DDDI-b gives a larger pairing gap at low and moderately low densities than
DDDI-a1 and DDDI-a2, while at densities around the saturation the gap becomes small 
and almost vanishing. \footnote{The auxiliary-field diffusion Monte Carlo method \cite{Gandolfi2008}, another ab initio
calculation, predicts the pairing gap close to the BCS gap in the density range $\rho_n \lesssim 10^{-2}$ $\mathrm{fm^{-3}}$.
Thus it may be considered that the parameter set DDDI-b represents partly this result.} 
We consider that  DDDI-a1 and DDDI-a2
are more realistic than DDDI-b while the difference between DDDI-a1 and DDDI-a2 
represents the uncertainty in modeling the realistic pairing correlation. We also use
the  model DDDI-b since it simulates the BCS gap, which is a robust baseline common
to all the models of realistic  bare nuclear force \cite{Dean2003}.

\begin{figure}
 \centering
 \includegraphics[width=0.6\hsize, bb = 0 0 300 200]{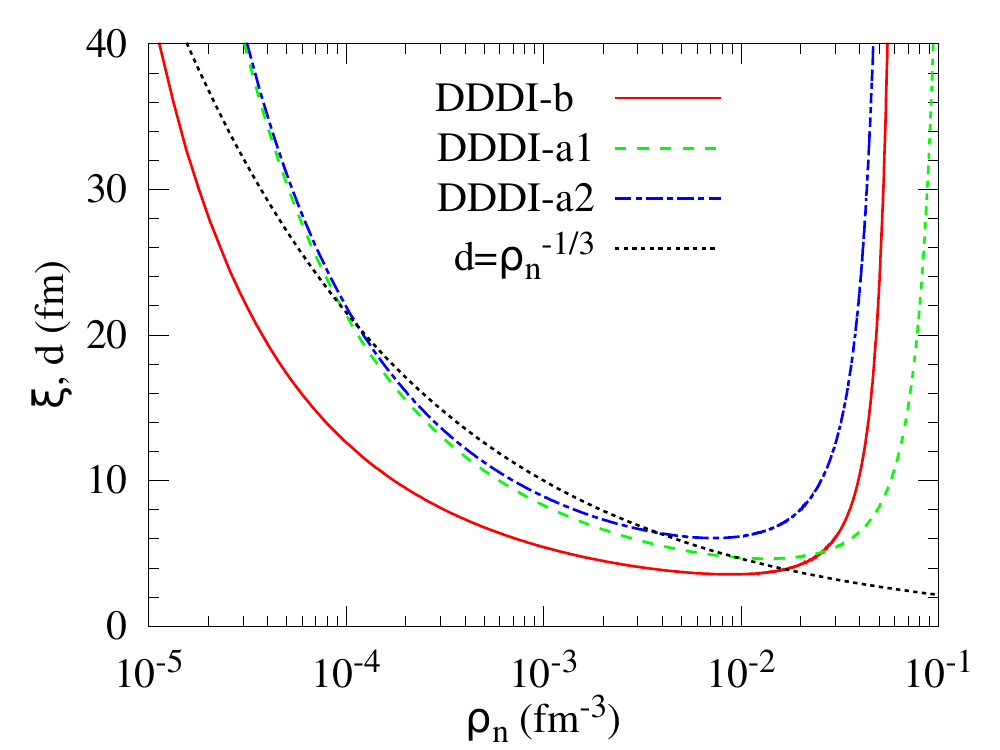}
    \caption{
    The coherence length $\xi$ of superfluid neutron matter obtained in the uniform-BCS calculation.
    The dotted curve is the average inter-neutron distance $d=\rho_n^{-1/3}$.
    See also the caption of Fig.~\ref{fig1}. }
  \label{fig2}    
 \end{figure}

Figure \ref{fig2} shows the coherence length $\xi$ of superfluid uniform neutron matter,
calculated as described in Appendix B. 
The coherence length $\xi$ depends strongly on the neutron density. The coherence length $\xi$ is as short
as $\xi \lesim 10$ fm at  
$\rho_n=10^{-3}- 2\times 10^{-2}$ fm$^{-3}$.  
The coherence length  becomes long gradually as
the neutron density decreases less than $10^{-3}$ fm$^{-3}$, and it also does rather sharply
for increasing $\rho_n$ more than  $\sim 3\times 10^{-2}$ fm$^{-3}$.  
The minimum value of the coherence length is $\xi\sim 3.6$ fm for DDDI-b at neutron density
corresponding to $\lambda_n \approx 5$ MeV, 
 $\xi\sim 4.6$ fm for DDDI-a1 at $\lambda_n \approx 6$ MeV, and
 $\xi\sim 6.1$ fm for DDDI-a2 at $\lambda_n \approx 5$ MeV.
 The dotted curve in Fig. \ref{fig2} shows the average inter-neutron distance $d=\rho_n^{-1/3}$.
  It is noted that
 the coherence length $\xi$ is shorter than the average inter-neutron distance $d=\rho_n^{-1/3}$
 at wide density interval $\rho_n=10^{-5}-10^{-2}$ fm$^{-3}$ for DDDI-b, or
 comparable with $d$  at $\rho_n=10^{-4}-10^{-2}$ fm$^{-3}$ for DDDI-a1 and DDDI-a2.
 The  coherence length shorter than $d$ implies that the pair correlation at these densities is
 in the domain of the strong-coupling pairing, characterized as 
  the BCS-BEC crossover phenomenon \cite{Strinati2018}.

\begin{figure}
 \centering
 \includegraphics[width=0.6\hsize, bb = 0 0 288 216]{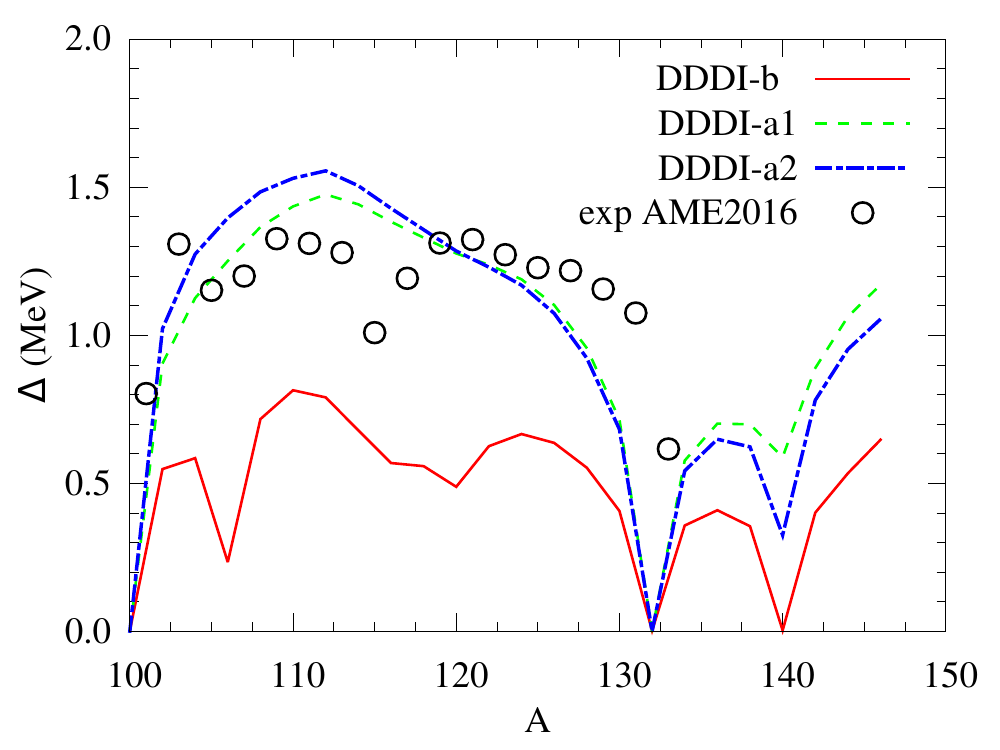}
    \caption{
    The neutron pair gap in Sn isotopes obtained with the Skyrme-Hartree-Fock-Bogoliubov
    method using the three DDDI pairing interaction models. 
    Solid, dashed and dot-dashed curves correspond to DDDI-b,
    DDDI-a1 and DDDI-a2, respectively. The Skyrme parameter SLy4 is
    adopted. The open circle is the experimental neutron pair gap derived using the
    odd-even mass difference \cite{Satula1998} and AME2016 \cite{AME2016}. See text for details.}
      \label{fig3}    
 \end{figure}

 Figure \ref{fig3} shows the average neutron pairing gap $\Delta_{n,uv}$ in Sn isotopes obtained in the present HFB code.
 The average neutron pairing gap in $^{120}$Sn calculated with DDDI-b, DDDI-a1 and
DDDI-a2  is $\Delta_{n,uv}=0.48, 1.28$ and 1.28 MeV, respectively. Note that the 
pair gap of DDDI-a1 and DDDI-a2 reproduces the experimental gap
reasonably well over the long isotope chain while DDDI-b gives only a
half of the experimental value. 
 
   \begin{figure}[t]
    \includegraphics[width=.8\hsize, bb= 0 0 288 216]{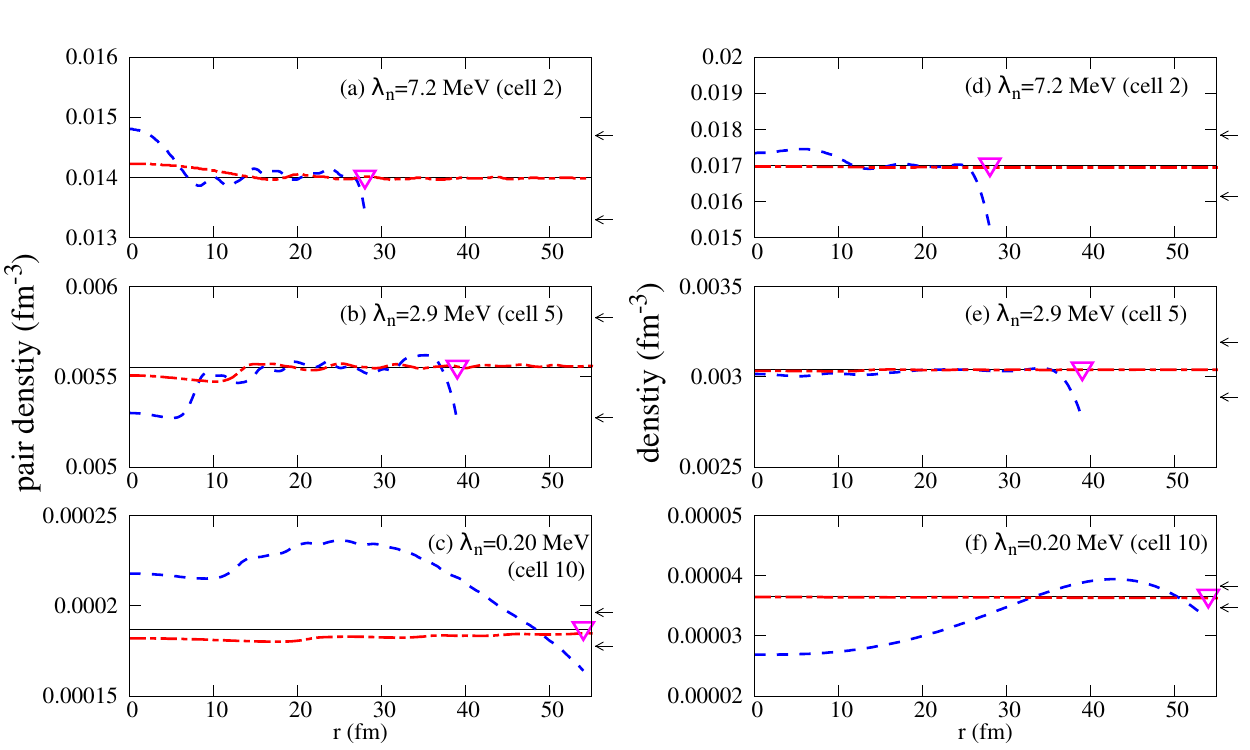}
   \caption{Panels (a)(b) and (c): The pair density $\tilde{\rho}_n(r)$
   of pure neutron matter obtained with the HFB calculation with a spherical box for the neutron
   Fermi energy; (a) $\lambda_n=7.2$ MeV,  (b) 2.9 MeV and (c) 0.20 MeV, which corresponds to cells 2,5 and 10 in
   Table. \ref{nvnuc}. The parameter set DDDI-a1 is used for the DDDI pairing interaction.
   The dashed curve represents the result of the calculation where the box size is chosen the same as
   the Wigner-Seitz radius $R_\mathrm{box}=R_\mathrm{cell}=28$ fm, 39 fm, and 54 fm
   of the respective cells  (indicated by the triangle symbol). The red dot-dashed
   curve is the result obtained with a large box size $R_\mathrm{box}=100$ fm for (a) and (b)
   and  $R_\mathrm{box}=200$ fm for (c). The horizontal line is the result of the
   uniform-BCS calculation obtained with the same Fermi energy. The arrows in the right vertical
   axis indicate the deviation of $\pm 5\%$ from the uniform-BCS result.
   Panels (d)(e) and (f): the same as (a)(b) and (c), but for
   the number density $\rho_n(r)$. }
   \label{fig4}
  \end{figure}

 \section{Finite-size effect and large-box configuration}

 Since the present HFB calculation is performed in the radial coordinate space
truncated with a finite box radius  $R_\mathrm{box}$, obtained results depend
on the box radius $R_\mathrm{box}$ especially when $R_\mathrm{box}$ is not large. This kind of dependence
is often called   
the finite-size effect. 
If we adopt the Wigner-Seitz approximation, where the box size is chosen equal to the Wigner-Seitz radius $R_\mathrm{cell}$ 
of the lattice cell of the inner crust, results also include the finite-size effect. 
We shall examine how the Wigner-Seitz approximation is affected by the finite size effect. For this purpose,
we here describe pure neutron matter using the same HFB code. For pure
 neutron matter,  we can obtain  an accurate numerical result by means of the uniform-BCS calculation,
 which corresponds to 
 the limit of
 infinite size $R_\mathrm{box} \rightarrow \infty$.
 Comparison with the uniform-BCS result makes it possible to evaluate the finite size effect. 
 
 We have applied the present HFB model to the pure neutron systems by simply neglecting the proton contributions. 
 Figure ~\ref{fig4} shows a 
 few example of the results, in which 
 the neutron Fermi energy is chosen as
 $\lambda_n=7.2$,  $2.9$ and 0.2 MeV corresponding to
   cells  2, 5 and 10 in Table \ref{nvnuc}, and
the neutron 
 density
  $1.8 \times 10^{-2}$,  $3.0\times 10^{-3}$ and $3.0 \times 10^{-4}$ fm$^{-3}$, 
 respectively.
 The pairing interaction DDDI-a1
 is adopted. 
 
 Dashed curves are the results for the calculation
 in which 
 the box radius $R_\mathrm{box}$ is set to the radius $R_\mathrm{cell}=28$, 39 and 54 fm of 
 the corresponding Wigner-Seitz cells. 
 It is seen that both the number density and the pair density of neutrons deviate from 
 the uniform-BCS results; the finite size effect in the pair density is not negligible and 
 much larger than that
for the number density. 
The deviation
from the uniform-BCS result (horizontal lines) is more than 20\% in 
cell 10 although it is less than about 5\% in the other cells 2 and 5.
The boundary condition with the finite box causes discretization of the energy
spectrum of the quasiparticle states, and the pairing property is influenced
by the discretization if the pair gap is not large enough than the energy spacing.  
It is also seen that the deviation from the uniform-BCS is
worse at positions close to the origin than at far positions. A possible explanation is
that   the influence of the
discretization of the quasiparticle energy spectrum may be stronger at small $r$
than at larger $r$; the number of contributing quasiparticle states
is effectively small since the wave function of
high-$\ell$ partial waves is suppressed at  small $r$.

The above results indicate that the Wigner-Seitz approximation to the inner crust matter
may not be accurate enough to discuss the proximity effect. One needs to control the finite
size effect in a better way.  A desirable approach may be to take into account the
lattice structure of the inner crust matter using the band theory method and the Bloch
waves, where the continuity of the neutron quasiparticle spectrum is kept. However the
band theory applied to the HFB calculation is presently quite limited \cite{chamel2010},
and a calculation with a large quasiparticle space is too demanding and difficult to be performed. Instead 
we adopt  a simpler 
approach where a nuclear cluster is placed in a neutron superfluid confined in
a large box, where the box size is chosen sufficiently large in order to reduce the finite-size effect
 as much as possible.  

We find that 
$R_\mathrm{box}\gesim 100$ fm gives the pair density convergent  to the uniform-BCS
with accuracy of around 1\% for densities $\rho_n \gesim 1 \times 10^{-4}$ fm$^{-3}$
as shown in   Figure \ref{fig2}(a)(b), where we plot the results obtained with 
$R_\mathrm{box}= 100$ fm. In very-low-density cases $\rho_n \lesim 1 \times 10^{-5}$ fm$^{-3}$,  
the pairing gap becomes very small $\Delta \lesim 0.01$ MeV. In this case, influence of the discretization
in quasiparticle levels is less negligible, and hence a larger box is required.   For cell 10 (Fig. \ref{fig2}(c)),  
we obtained the agreement to the required accuracy with $R_\mathrm{box}= 200$ fm.

In the following we adopt this large-box configuration to discuss the proximity effect associated with the
presence of the nuclear cluster.

 \section{Proximity effect}
 
     \begin{figure}[tp]
      \centering
     \includegraphics[width=.5\hsize, bb= 0 0 300 200]{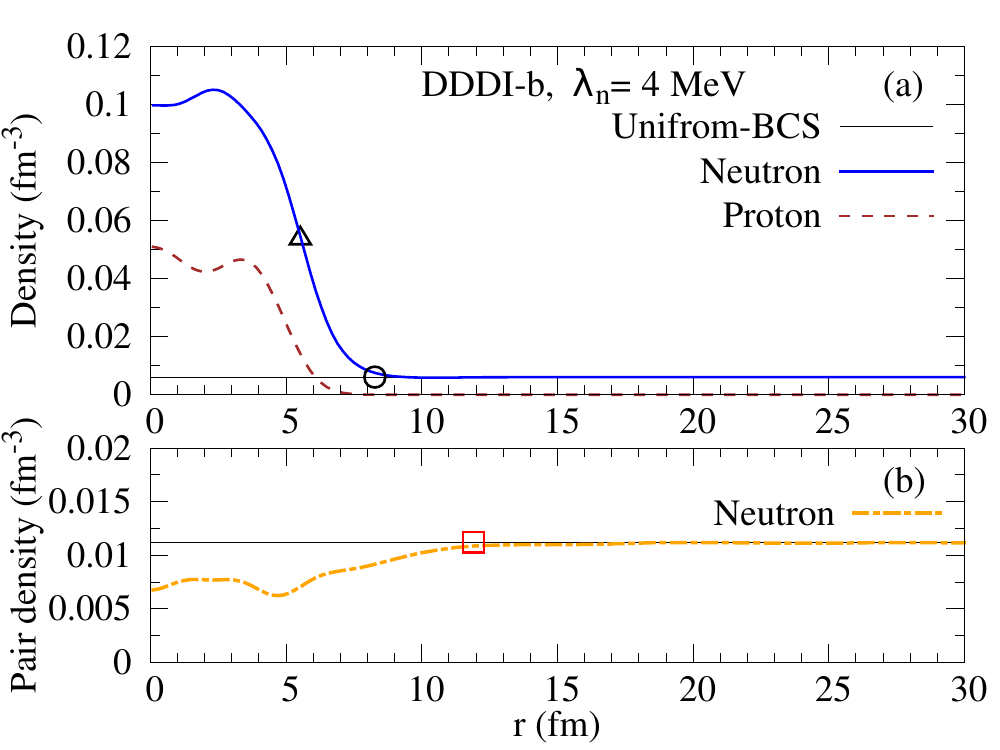}
          \caption{Result of the HFB calculation for the configuration with the proton number $Z=28$, 
          the neutron Fermi energy $\lambda_n=4$ MeV, and the box size $R_\mathrm{box}=100$ fm. The parameter set DDDI-b is
          used. 
          Panel (a) shows the number densities $\rho_n(r)$ and $\rho_p(r)$ of neutrons and protons, respectively.
          Panel (b) shows the neutron pair density $\tilde{\rho}_n(r)$.
          The horizontal lines are the uniform-BCS results with the same $\lambda_n$. The triangle and circle symbols
          indicate the half-density surface $R_\mathrm{s}$ and the edge radius $R_\mathrm{edge}$ of the cluster
          whereas the square symbol points to $R_\mathrm{edge}+\xi$. 
          See the text for the definitions of these quantities.
                   }
         \label{fig5}
     \end{figure}
     
We shall now discuss the pair correlation in the inner crust matter.
 As discussed above we consider the system confined in a large box, at the center of which  
a nuclear cluster is placed. Using this setup, we shall 
investigate how the presence of the nuclear cluster influences
the pair correlation of neutron superfluid in the neighborhood region around the cluster.

\subsection{length of the proximity effect}

In order to investigate 
general features of the proximity effect,
we shall first examine cases where the density of the surrounding neutron superfluid
is systematically varied while the proton number is fixed.
 In the next subsection we discuss realistic configurations of the inner crust matter, for which the proton number
and the density of neutron superfluid are chosen to represent
various layers of the inner crust.

The proton number is $Z=28$ in all the examples in this subsection and we vary the
neutron Fermi energy $\lambda_n$ systematically from 0.2 MeV to 6 MeV, which
corresponds to 
the 
density of the uniform neutron superfluid  from 
$\rho_{n}= 4\times 10^{-5}$ fm$^{-3}$ to $1 \times 10^{-2}$ fm$^{-3}$.  

A typical
result obtained for $\lambda_n = 4$ MeV ($\rho_{n}=6.1 \times 10^{-3}$ fm$^{-3}$) with DDDI-b is shown 
in Fig.~\ref{fig5}, where plotted are the number densities of neutrons and protons, $\rho_n(r)$ and
$\rho_p(r)$,  and the neutron
pair density $\tilde{\rho}_n(r)$ as a function of the radial coordinate $r$.
It is seen that the nuclear cluster is well localized in a central region as
seen in the profile of the neutron density $\rho_n(r)$ 
which converges rather quickly to a constant value at around $r \approx 8$ fm
(the proton density $\rho_p(r)$ converges to zero around $r \approx 6$ fm. ).
The surface of the nuclear cluster may be quantified by fitting to the neutron density with
a function of the Woods-Saxon type, 
  \begin{align}
 f_\mathrm{ws}(r)=\rho_{n,\mathrm{M}}+\frac{f_0}{1+\exp\left(\frac{r-R_\mathrm{s}}{a}\right)},
 \label{ws}
\end{align}
where $R_\mathrm{s}$ defines the half-density surface, and $a$ represents the diffuseness of the surface. 
The constant
$ \rho_{n,\mathrm{M}}$ is the neutron density obtained from the uniform-BCS performed
 for the same value of $\lambda_n$. The values of $f_0$, $R_\mathrm{s}$ and $a$ are extracted from a fitting.
 In addition we find it useful to consider
``the edge" of the nuclear cluster to evaluate the area where the cluster exists.
 We define the nuclear edge by
$R_\mathrm{edge}=R_\mathrm{s}+4a$. 
The  edge position 
$r=R_\mathrm{edge}$ is indicated by the black circle in Fig. \ref{fig5}, and it is seen
that $R_\mathrm{edge}$ represents well the position where the neutron density $\rho_n(r)$ 
converges to $\rho_{n,\mathrm{M}}$.  
 
 A most noticeable feature in Fig. \ref{fig5} is that 
 the neutron pair density $\tilde{\rho}_n(r)$  exhibits behaviours different from those of  
 the neutron number density $\rho_n(r)$. It is
 seen that the neutron pair density $\tilde{\rho}_n(r)$ slowly converges and reaches the uniform-BCS value at around $r\approx 12$ fm,
  deviating from $R_\mathrm{edge}$ by about 4 fm.
In other words the influence of the nuclear cluster extends to the neighbour region beyond $R_\mathrm{edge}$. 
 This slow convergence is nothing but the proximity effect. 
 In this example the neutron pair density inside the
 cluster is significantly smaller than that outside the cluster. This reflects the characteristic 
 density dependence of the neutron pair gap of the DDDI-b model; 
 the gap for  the density inside
 the cluster ($\rho_n \sim \rho_0$) 
 is very small $\Delta \lesim 0.1$ MeV whereas that 
 for the density of neutron superfluid 
  ($\rho_{n,\mathrm{M}} \sim 6.1 \times 10^{-3}$ fm) is relatively large $\Delta \sim 2.1$ MeV.
 
 It has been argued that the proximity effect emerges in a region adjacent to the border 
 with its length scale characterized by the coherence length
 $\xi$ of the superfluid/superconducting matter \cite{gennes1964}.
 We here assume that the border between the neutron superfluid and the nuclear cluster is approximated by the edge radius $R_\mathrm{edge}$, rather than the half-density
 surface $R_\mathrm{s}$. If these considerations are reasonable, it is expected that 
 the proximity effect is seen up to $r \approx 
R_\mathrm{edge}+\xi$. 
 In the case shown in Fig.~\ref{fig5}, the position 
 where  the neutron pair density converges to the uniform-BCS value 
 corresponds well to $r = 
R_\mathrm{edge}+\xi = 8.27$ fm$+3.63$ fm $=$ 11.9 fm, and the above argument appears to hold.

\begin{figure}[tp]
\centering
       \includegraphics[width=\hsize, bb= 0 0 576 432]{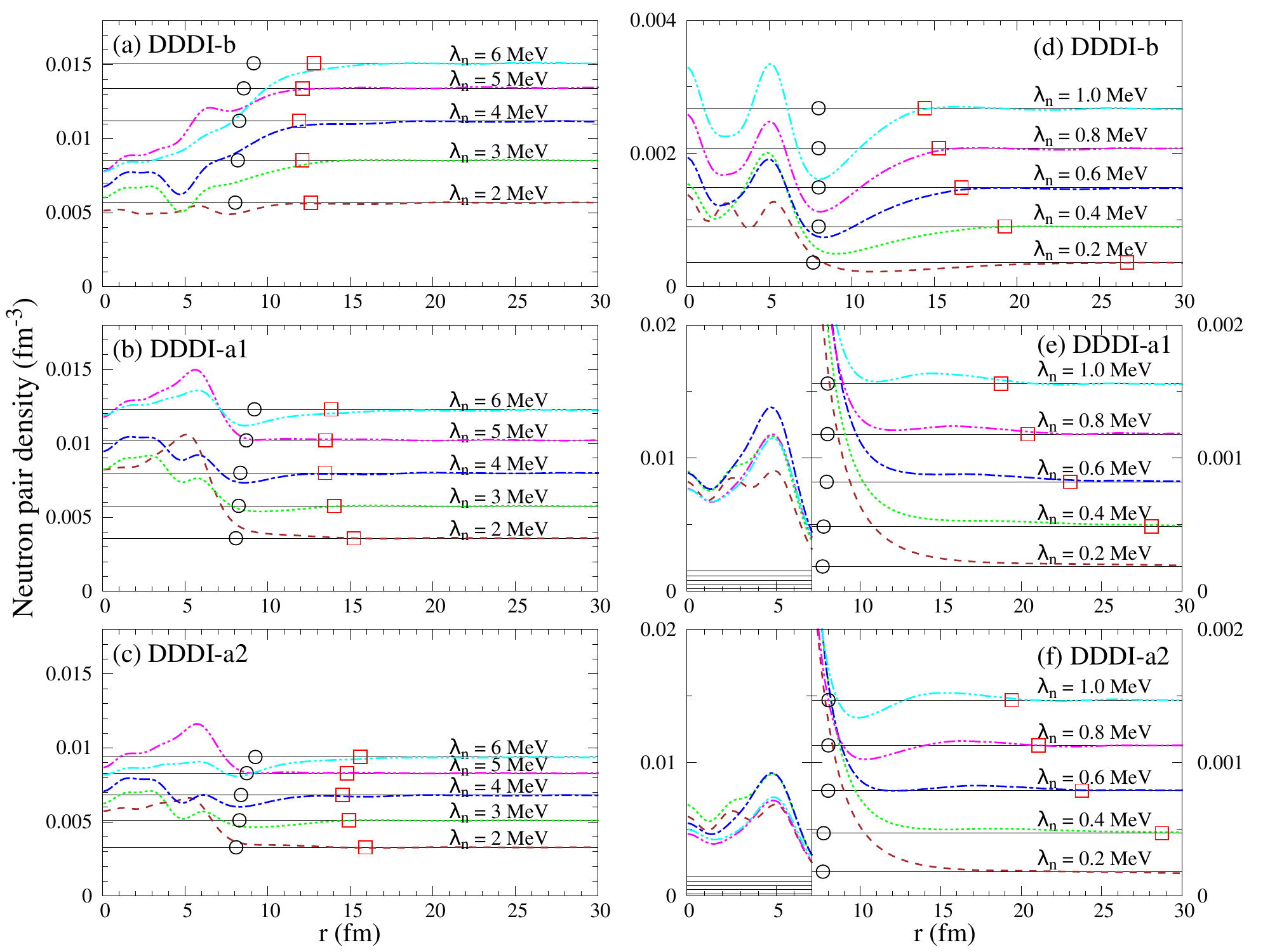}
       \caption{Calculated neutron pair density $\tilde{\rho}_n(r)$ for the neutron Fermi energy $\lambda_n=0.1-1.0$ MeV
       and 2-6 MeV with the three DDDI parameter sets.
       Panels (a)(d), (b)(e) and (c)(f) are for DDDI-b, DDDI-a1 and DDDI-a2, respectively.
       The horizontal line is the results of the uniform-BCS calculation.
       For the symbols, see the caption of Fig. \ref{fig5}.}
 \label{fig6}   
     \end{figure}

 Figure \ref{fig6} show systematic behaviours of the neutron pair densities calculated for various neutron Fermi energies 
 and for three different pairing interactions: the DDDI-b (panel (a) in each figure), the DDDI-a1 (b), and 
 the DDDI-a2 (c). Figure \ref{fig6}(a)(b)(c)  shows the results for
 the neutron Fermi energy $\lambda_n=2-6$ MeV corresponding
 the neutron density $\rho_n \sim 10^{-3}-10^{-2}$ fm$^{-3}$ (see Fig. \ref{fig1}),
 and Fig. \ref{fig6}(d)(e)(f)  for  $\lambda_n=0.2-1\,\mathrm{MeV}$ 
 ($\rho_n \sim 10^{-4}-10^{-3}$ fm$^{-3}$).

 The proximity effect is clearly visible in all the cases;  
 the pair density converges to
   that of the uniform neutron superfluid at a position deviating significantly
   from the edge  position $r=R_\mathrm{edge}$ of the nuclear cluster.
   It is also seen that the range of the proximity effect depends rather strongly
on the neutron Fermi energy or the density of the neutron superfluid,
especially at { low neutron density $\rho_{n,\mathrm{M}} \lesim 5\times 10^{-4}$ fm$^{-3}$
and $\lambda_n \lesim 1.0$ MeV.
It also depends on the three DDDI models. Despite the differences in the pairing properties,
we confirm here that 
the range where the proximity effect reaches} is described well
by the position $r = R_\mathrm{edge} + \xi$ (marked with the square symbol), 
characterized by the coherence length $\xi$ measured from the edge $R_\mathrm{edge}$
of the nuclear cluster. 
(Note that the edge position $R_\mathrm{edge}$ 
 of the nuclear cluster depends only weakly on the neutron Fermi energy, and there is
 essentially no dependence on the three choices  of the pairing interaction. )

We here recall Fig.~\ref{fig2} where 
the coherence length is shown to become as small as $\lesim 10$ fm at moderately low
density $\rho_n=7\times 10^{-4}-2\times 10^{-2}$ fm$^{-3}$ for the three DDDI's.
This brings about the short range of  the proximity effect seen for 
$\lambda_n=2-6$ MeV. 
 This is related to the specific feature of
the dilute neutron superfluid that  the BCS-BEC crossover is about to occur at these densities.

A long range of the proximity effect seen for 
$\lambda_n=0.2-1.0$ MeV can be related to  the monotonic and considerable
increase of the coherence length $\xi$ with decreasing neutron density
for very low density $\rho_n \lesim 10^{-3}$ fm$^{-3}$. Note that
for $\rho_n\sim10^{-5}-10^{-4}$ fm$^{-3}$, the coherence length
$\xi=5-20$ fm in the case of DDDI-b, $\xi=10-36$ fm for DDDI-a1 and $\xi=11-37$ fm for DDDI-a2.
If the density of the external neutron superfluid decreases further, 
the range of the proximity effect is expected to exceed far beyond 50 fm.

        \subsection{Realistic inner crust configurations}

          \begin{table}[t]
      \flushright
  \caption{The proton number $Z$ and the neutron Fermi energy $\lambda_n$ employed in the
  present calculation to represent realistic configurations of the inner crust cells \cite{NV1973}.
  The next columns are the density $\rho_{n,\mathrm{M}}$, the
  pairing gap $\Delta$  and the coherence length $\xi$, obtained from the uniform-BCS calculation for the corresponding 
  neutron matter.  
  Results of the three DDDI models,
  DDDI-b, DDDI-a1, and DDDI-a2, are listed for $\Delta$ and $\xi$ while $\rho_{n,\mathrm{M}}$ is shown
  only for DDDI-a1. The third last column is the edge radius  $R_\mathrm{edge}$ of 
  the nuclear cluster extracted from the Hartree-Fock-Bogoliubov calculation with DDDI-a1.
  The second last is the Wigner-Seitz radius $R_\mathrm{cell}$ of the cells\cite{NV1973} 
  while the last is the average baryon density $\rho_b=\int_0^{R_\mathrm{cell}}(\rho_n(r)+\rho_p(r))r^2 dr/(R_\mathrm{cell}^3/3) $ 
  (with DDDI-a1) evaluated using the same Wigner-Seitz radius.
  See text for details.} 
    \footnotesize
     \begin{minipage}{\hsize}
   \begin{tabular}[t]{crcccccccccccccccc}
    \hline
    Cell&$Z$  &$\lambda_n$& $\rho_{n,\mathrm{M}}^\mathrm{a1}$ &
    $\Delta^\mathrm{b} $&$\Delta^\mathrm{a1} $&$\Delta^\mathrm{a2}$ &
    $\xi^\mathrm{b} $&$\xi^\mathrm{a1}$&$\xi^\mathrm{a2}$&
    $R^\mathrm{a1}_\mathrm{edge}$ & $R_\mathrm{cell}$ & $\rho_\mathrm{b}^\mathrm{a1}$  \\
    &&(MeV)&($\mathrm{fm^{-3}}$)&(MeV)&(MeV)&(MeV)&(fm)&(fm)&(fm)&(fm)&(fm)& $(\mathrm{fm^{-3}})$\\\hline
1&$\mathrm{Zr}$ &11.00 &$4.44\times 10^{-2}$ &0.83&1.66&0.36 &14.09&7.17&31.80 &12.90  &20 & $5.02 \times 10^{-2}$ \\ 
2&$\mathrm{Sn}$ &7.20 &$1.80\times 10^{-2}$  &2.40&2.03&1.26 &4.06&4.71&7.33 &11.01 &28 & $2.04 \times 10^{-2}$\\
3&$\mathrm{Sn}$ &4.80 &$7.81\times 10^{-3}$  &2.31&1.60&1.25 &3.56&4.84&6.06 &10.12 &33   & $9.19 \times 10^{-3}$\\ 
4&$\mathrm{Sn}$ &3.70 &$4.75\times 10^{-3}$  &1.98&1.28&1.06 &3.69&5.30&6.24 &9.72  &36 & $5.69 \times 10^{-3}$\\ 
5&$\mathrm{Sn}$ &2.90 &$3.04\times 10^{-3}$  &1.64&1.01&0.88 &3.96&5.90&6.71 &9.40  &39 & $3.75\times 10^{-3}$\\ 
6&$\mathrm{Zr}$ &1.70 &$1.19\times 10^{-3}$  &1.01&0.58&0.53 &4.91&7.82&8.48 &9.00 &42 & $1.63\times 10^{-3}$\\ 
7&$\mathrm{Zr}$ &1.00 &$4.90\times 10^{-4}$  &0.59&0.32&0.30 &6.42&10.77&11.38 &8.78 &44 & $8.59\times 10^{-4}$\\ 
8&$\mathrm{Zr}$ &0.70 &$2.73\times 10^{-4}$  &0.40&0.21&0.20 &7.88&13.62&14.23 &8.79 &46 & $5.96\times 10^{-4}$\\ 
9&$\mathrm{Zr}$ &0.37 &$9.74\times 10^{-5}$  &0.18&0.09&0.09 &11.90&21.63&22.30 &8.62 &49 & $3.62\times 10^{-4}$\\ 
10&$\mathrm{Zr}$ &0.20 &$3.65\times 10^{-5}$ &0.08&0.04&0.04 &19.01&36.09&36.87 &8.46 &54 & $2.30\times 10^{-4}$\\\hline 
   \end{tabular}
       \label{nvnuc}     
            \end{minipage}
  \end{table}
  
    \begin{figure}[t]
   \centering
   \includegraphics[width=.6\hsize, bb= 0 0 300 200]{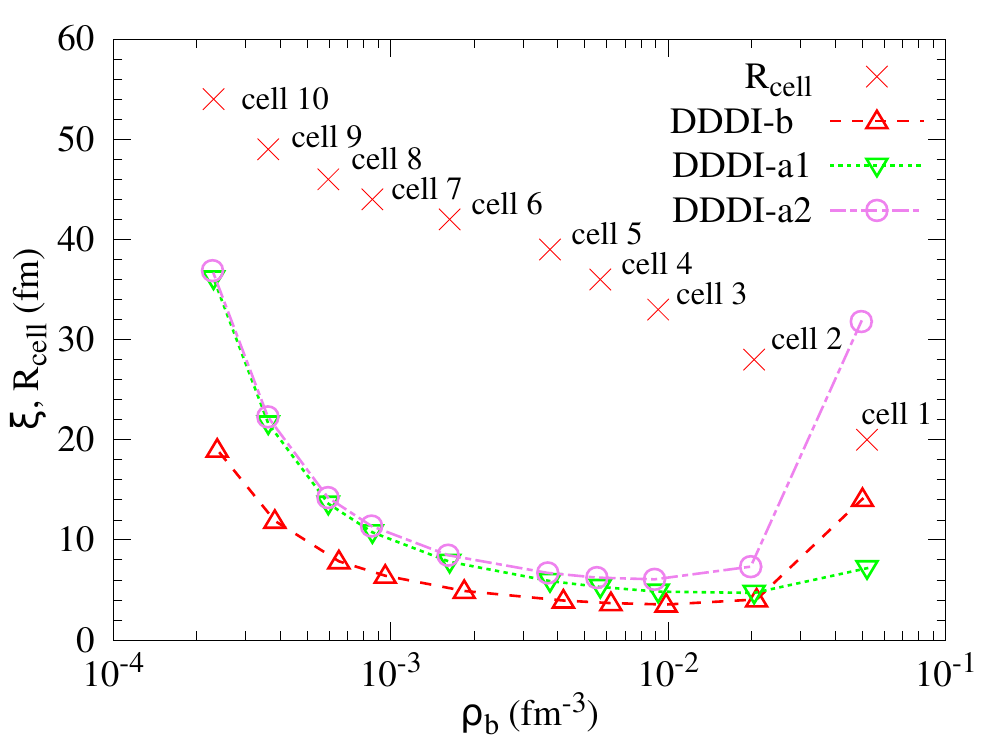}
 \caption{The coherence length $\xi$ of uniform neutron superfluid corresponding to  cells 1 to 10  listed in Table \ref{nvnuc}.
 Results of the three gap models, DDDI-b, DDDI-a1 and DDDI-a2, are shown. 
 The horizontal axis is the baryon density $\rho_b$ of the cells.  
 The Wigner-Seitz radius $R_\mathrm{cell}$ of the cells, taken from Ref. \cite{NV1973}, is also plotted for
 comparison.}
  \label{xi}
  \end{figure}

    \begin{figure}[tb]
     \centering
  \includegraphics[width=\hsize, bb= 0 0 360 216]{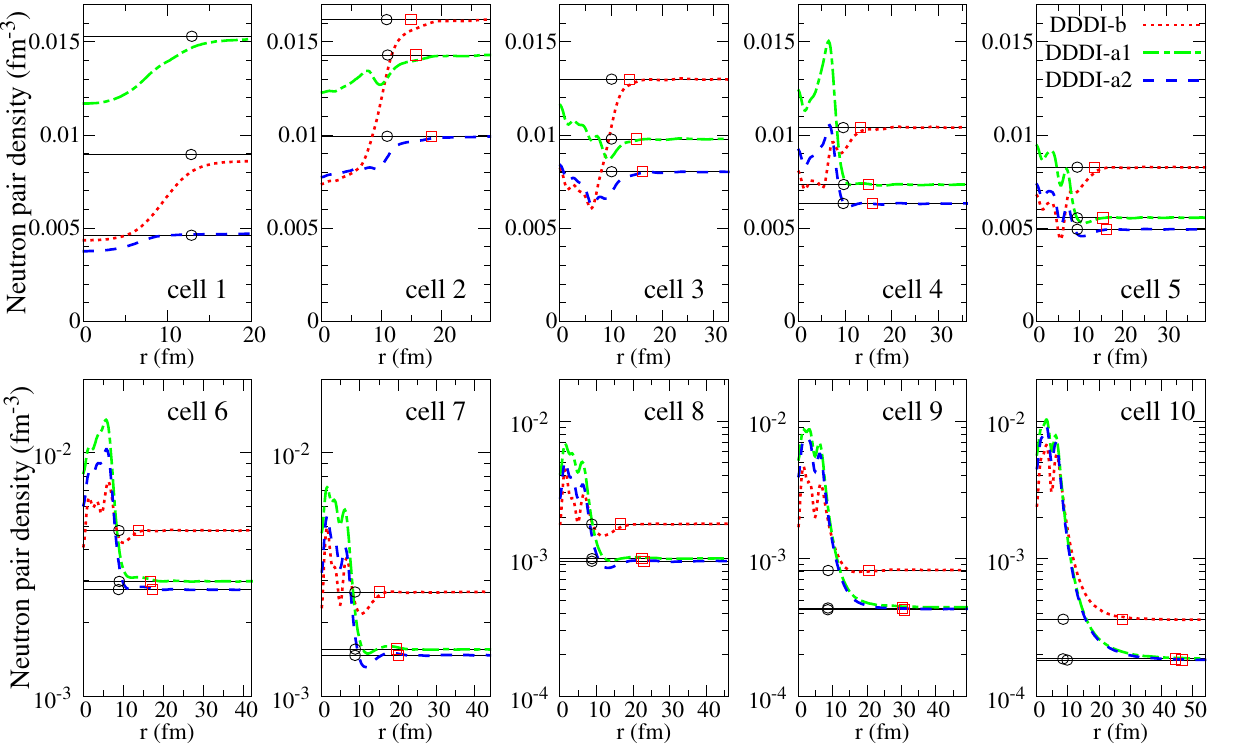}
  \caption{Calculated neutron pair densities in various cells in the inner crust of neutron stars,
  listed in Table \ref{nvnuc}, obtained with 
  three DDDI models, DDDI-b (dotted curve), DDDI-a1 (dot-dashed), and DDDI-a2 (dashed). 
         The horizontal line is the results of the uniform-BCS calculation. For the symbols, see the caption of Fig. \ref{fig5}. }
  \label{nvrho}
    \end{figure}
    
    Finally, we discuss the proximity effect for realistic situations of 
    the inner crust of neutron stars. Here we refer to the Wigner-Seitz cells obtained in Negele and Vauthrin \cite{NV1973}
    for various layers of the inner crust.  We perform the HFB calculation 
    for the cells listed in
   Table \ref{nvnuc} using the large-box configuration.  The proton number $Z$ and the
   Wigner-Seitz radius $R_\mathrm{cell}$ of each cell is taken from
   Ref. \cite{NV1973}.  The neutron Fermi energy $\lambda_n$, the control parameter
   of the neutron density, is chosen 
    so that the obtained density of the external neutron superfluid
   reproduces approximately the density of the neutron gas in Ref. \cite{NV1973}. For simplicity
   we use a common value of $\lambda_n$ for the three DDDI models. The box size is
   $R_\mathrm{box}=100$ fm for most cells and 200 fm only for cell 1 with DDDI-b2, cell 10 with DDDI-a1 and cell 10 with DDDI-a2.

   The calculated neutron pair density  is shown in Fig. \ref{nvrho}. The maximum of the plotted radial
   coordinate  is the Wigner-Seitz radius $R_\mathrm{cell}$ for each cell.
 A noticeable feature is that  in cells  3 to 8
the pair density converges to that of the uniform-BCS at  a distance shorter than  the half distance of the Wigner-Seitz radius. In other words the proximity effect is restricted only in a small area nearby the
 nuclear cluster. The area of uniform neutron superfluid  and that of the nuclear cluster are
 well separated  in  these middle layers of the inner crust. 
 This feature is common to the three  DDDI pairing models.
It is noted that the coherence length of external neutron superfluid 
 is the smallest $\xi \approx 4-6$ fm at cells 2-6 (for DDDI-b), cells 2-5 (for DDDI-a1), and 
 cell 3 (for DDDI-a2), which are significantly smaller than the Wigner-Seitz radius of these
 cells at  the middle layers.
  
In cells 9 and 10, where the external neutron superfluid is
dilute ($\rho_{n,\mathrm{M}} \lesim 1 \times 10^{-4}$ fm$^{-3}$), 
the proximity effect extends to a major area of
 the Wigner-Seitz cell, beyond the half length of
 the Wigner-Seitz radius, especially for DDDI-a1 and DDDI-a2.  This reflects 
the long coherence length at such very low densities: $\xi \gesim 20-40$ fm
 for DDDI-a1 and DDDI-a2,  and  $\xi \gesim 12-20$ fm for DDDI-b.
 Note that the pairing gap of DDDI-a1 and DDDI-a2 
 in dilute neutron matter is reduced from 
 the BCS value (corresponding to DDDI-b) by a factor of about 2, 
 leading to a longer coherence length in these realistic gap models DDDI-a1 and DDDI-a2.
 
Another case where a long-range proximity effect is predicted is  cell 1 at relatively high density,
where the external neutron density $\rho_{n,\mathrm{M}} \sim 0.04$ fm$^{-3}\approx \rho_0/2$
is about a half of that of the saturated nuclear matter. 
 The pair density deviates from that of the uniform neutron superfluid in the whole area of
 the Wigner-Seitz cell. In this cell with relatively high neutron density 
 the predicted coherence length $\xi$ varies from 7 to 30 fm depending rather strongly on the pairing models, reflecting the uncertainty of the gap at such density.
 However, because of the relatively high baryon density and  a large N/Z ratio, 
  the Wigner-Seitz radius $R_\mathrm{cell}$ becomes small ($\sim 20$ fm) and
  the edge position $r=R_\mathrm{edge}$ of the nuclear cluster becomes as large as $\sim 13$ fm
  due to a thick neutron skin of the cluster.
Consequently the range $R_\mathrm{edge} +\xi$ of the proximity effect
 exceeds the Wigner-Seitz radius
irrespective of the uncertainty of the pairing gap.
  Note  that  cell 1 corresponds to a deep layer of the inner crust, where a transition to the
  so called  pasta phase is about to occur. The present result suggests 
 strong proximity effect also for the pasta phase at higher baryon density.    
 We remark also that the
 proximity effect in these deep layers might be even stronger than the present 
 prediction because of the presence of adjacent nuclear clusters in  the lattice configuration,
 but a quantitative evaluation is beyond the scope of the present study.

\section{Conclusion}
We have studied in detail 
the proximity effect of neutron pair correlation in the inner crust of neutron stars by
applying the Skyrme-Hartree-Fock-Bogoliubov theory formulated in the
coordinate representation. We describe a many-nucleon system consisting of $Z$ protons 
 (which form a nuclear cluster) and neutrons 
with a given positive Fermi energy, confined in a spherical box.
If we choose the box radius $R_\mathrm{box}$ equal
to the Wigner-Seitz radius of the lattice cell, the calculation
corresponds to the Wigner-Seitz approximation often adopted in preceding studies.
We found however that for the realistic Wigner-Seitz radius $R_\mathrm{cell} \sim 20-50$ fm
of the inner crust matter, influence of the box truncation or the finite-size effect is not
negligible for quantitative analysis of the proximity effect. We therefore use a large-box
configuration where the box size is chosen sufficiently large $R_\mathrm{box} \geq 100$ fm.
In other words, we
considered a simplified model of the inner crust matter in which
a single nuclear cluster is immersed in a uniform neutron superfluid, prepared in a sufficiently
large box. 
As the effective interaction causing the pairing correlation, we introduced
new parameterizations of the density-dependent delta interaction (DDDI-a1 and DDDI-a2) so that they
reproduce the ab initio evaluations of the pair gap in low-density neutron matter
as well as the experimental pair gap in finite nuclei. 

Focusing on the neutron pair density $\tilde{\rho}_n(r)$ (i.e. a locally defined pair condensate), we have
examined how $\tilde{\rho}_n(r)$ is affected by the presence of the nuclear cluster and how 
this quantity around the cluster converges
to the limiting value of the immersing neutron superfluid.
It is found from a systematic analysis 
that range of the proximity effect is characterized by the
coherence length of neutron superfluid measured from the edge position of the 
cluster. An important feature is that the coherence length $\xi$ depends strongly on the
density $\rho_n$ of neutron superfluid. The coherence length is as short as $\xi \sim 5-8$ fm
for density $1 \times 10^{-3}$ fm$^{-3} \lesim \rho_n \lesim  2 \times 10^{-2}$ fm$^{-3}$ while
it increases gradually at lower density $\rho_n \lesim 1 \times 10^{-3}$ fm$^{-3}$
and rather quickly at higher density $\rho_n \gesim 3 \times 10^{-2}$ fm$^{-3}$.

Applying the above result to the realistic configurations of the inner crust, we predict that
the proximity effect
is well limited in the vicinity of the nuclear cluster, i.e. in a sufficiently smaller area than
the Wigner-Seitz cell in the middle layers of the inner crust with
baryon density $5 \times 10^{-4}$ fm$^{-3} \lesim \rho_b \lesim 2\times 10^{-2}$ fm$^{-3}$.
On the contrary, the proximity effect is predicted to
extend to the whole volume of the Wigner-Seitz cell in the shallow layers of
the inner crust with $\rho_b \lesim 2 \times 10^{-4}$ fm$^{-3}$. Another
region where the range of  the proximity effect is expected to cover the whole Wigner-Seitz cell is deep
layers of the inner crust with $\rho_b \gesim 5 \times 10^{-2}$ fm$^{-3}$,
where the Wigner-Seitz radius becomes small $R_\mathrm{cell} \lesim 20$ fm while the
coherence length may becomes comparable or larger than $R_\mathrm{cell}$. 
This observation indicates that in these layers there is no clear separation between the nuclear cluster
and the immersing neutron superfluid as far as the pairing correlation is concerned.
It implies that the phenomena originating from the pair correlation and superfluidity,
such as the vortex pinning and the superfluid phonon excitations
may also be affected by the proximity effect. It is noted also that theoretical
approaches taking into account the lattice configuration is preferred for such cases.
It is a subject to be pursued in future study. 

\section*{Acknowledgement}
We thank  T. Inakura, K. Sekizawa, and K. Yoshida  for valuable discussions.
We also thank A. Ohnishi for a critical comment on the DDDI models.
This work was supported by the JSPS KAKENHI (Grants Nos. 17K05436 and 20K03945).

 \section*{Appendix A: Effective contact interaction for the GMB gap}
 
Here we discuss the parameter set of DDDI which reproduces the pairing gap of
Gor'kov Melik-Barkuhudarov (GMB) in the dilute limit of neutron matter. 
This is introduced by combining the known arguments on the 
GMB pairing gap \cite{heiselberg2000,Pethick-Smith} and 
on the effective strength of the contact interaction \cite{Bertsch-Esbensen1991,Garrido1999}.  

Let us first outline the relation between the strength of the contact interaction and 
the pairing gap in the BCS approximation.
For the pairing interaction of the contact two-body force 
$v(\vec{r}_1-\vec{r}_2) = V_0 \delta (\vec{r}_1-\vec{r}_2)$ ,
the gap equation in the weak-coupling BCS approximation reads
\begin{equation}
\frac{1}{V_0}=-\frac{1}{2}\sum_{\vec{k}}\frac{1}{\sqrt{(e_k - \lambda)^2 + \Delta^2}}
\label{gapeq-bcs}
\end{equation}
where $e_k =\frac{\hbar^2k^2}{2m}$, $\lambda=e_F=\frac{\hbar^2k_F^2}{2m}$, and $\Delta$ is the single-particle energy, the Fermi energy
(with the Fermi momentum $k_F$) and the pairing gap, respectively. 
To avoid the divergence inherent to the contact interaction, 
the sum $\sum_{\vec{k}}\equiv \frac{1}{(2\pi)^3} \int_{0}^{k_\mathrm{cut}} 4\pi k^2 dk$ 
 is performed with a cut-off momentum $k_c$ or a cut-off single-particle energy 
 $e_\mathrm{cut} = \hbar^2 k_\mathrm{cut}^2/2m$. The
 force  strength $v_0$ can be chosen so that the same interaction reproduces the zero-energy T-matrix 
 $T_0 = \frac{4\pi\hbar^2 a}{m}$,
 and the scattering length $a$ of the nucleon scattering in the $^{1}S_{0}$ channel.  
 This requirement is expressed in terms of
 the Lippmann-Schwinger equation for the T-matrix, which can be written as
 \begin{equation}
\frac{1}{V_0} = \frac{1}{T_0} + \frac{1}{2}\sum_{\vec{k}}\frac{1}{e_k},
\label{tmatrix}
 \end{equation}
 which determines the force strength $V_0$ as \cite{Bertsch-Esbensen1991,Garrido1999}
 \begin{equation}
 V_0 = -\frac{2\pi^2\hbar^2}{m}\frac{1}{k_\mathrm{cut} - \frac{\pi}{2a}}.
 \end{equation}
 The gap equation (\ref{gapeq-bcs}) combined with the T-matrix equation (\ref{tmatrix}) is written as
 \begin{equation}
 \frac{1}{T_0} = -\frac{1}{2}\sum_{\vec{k}}\left( 
 \frac{1}{\sqrt{(e_k - e_F)^2 + \Delta^2}} - \frac{1}{e_k}
 \right).
 \label{gapeq-reg}
 \end{equation}

 The gap equation (\ref{gapeq-reg}) is known to be solved analytically in the low-density limit $k_F \rightarrow 0$
 satisfying $k_F |a| \ll 1$ and $k_F \ll k_c$ \cite{Marini1998,Papenbrock1999}. The right hand side of Eq. (\ref{gapeq-reg}) is evaluated as
 $\simeq N_0 \log\left( \frac{e^2 \Delta }{8e_F}\right)$, where $N_0=\frac{mk_F}{2\pi^2\hbar^2}$ 
 is the single-particle level density at the Fermi energy.    
 The paring gap in this limit is then  given \cite{Marini1998,Papenbrock1999,Pethick-Smith,Strinati2018} as 
 \begin{equation}
 \Delta_\mathrm{BCS} = \frac{8e_F}{e^2} \exp \left( \frac{1}{T_0N_0} \right) = \frac{8e_F}{e^2} \exp \left( \frac{\pi}{2k_F a}\right).
 \end{equation}
Note that the T-matrix $T_0$ plays a role of a renormalized interaction strength of the contact force. 
 
 It is known that the medium effect in the low-density limit can be evaluated perturbatively as originally discussed
 by Gor'kov and  Melik-Barkhudarov \cite{gmb}. The effect is represented as an induced interaction \cite{heiselberg2000,Pethick-Smith}
 $U_\mathrm{ind}=N_0 T_0^2 (1+2\log 2)/3$ which modifies the interaction  strength 
 $T_0 \rightarrow T_0 + U_\mathrm{ind}$,
 where the numerical factor $(1+2\log 2)/3$ arises from an average of the Lindhard function.
 Similarly the left hand side of the gap equation (\ref{gapeq-reg}) is modified as
 \begin{equation}
 \frac{1}{T_0} \rightarrow \frac{1}{T_0 + U_\mathrm{ind}} \simeq \frac{1}{T_0} - \frac{1+2\log 2}{3}N_0,
 \label{induced}
 \end{equation}
 and hence the GMB pairing gap $\Delta_\mathrm{GMB}$ valid in the low-density limit is  given as
 \begin{equation}
 \Delta_\mathrm{GMB} = \frac{8e_F}{e^2} \exp \left( \frac{1}{T_0N_0}- \frac{1+2\log 2}{3}\right) 
 = \frac{1}{(4e)^{1/3}}\Delta_\mathrm{BCS}
 \end{equation} 
 with a reduction of a factor of $\simeq 1/2.2$ from the BCS gap.
 
 Now, by combining the argument on the contact force,
 Eq. (\ref{tmatrix}), and on the induced interaction modifying the l.h.s of the gap equation, Eq. (\ref{induced}),
 we find that an effective strength $V_\mathrm{GMB}$ of the contact force which reproduces the GMB pairing
 gap is given by 
 \begin{equation}
 \frac{1}{V_\mathrm{GMB}}= \frac{1}{T_0} -  \frac{1+2\log 2}{3} N_0+ \frac{1}{2}\sum_{\vec{k}}\frac{1}{e_k},
 \end{equation}
 which determines $V_\mathrm{GMB}$ as
 \begin{equation}
 V_\mathrm{GMB}=V_0\left\{ 1 - \frac{1+2\log 2}{3}\frac{k_F}{k_\mathrm{cut} - \frac{\pi}{2a}}
 + \mathrm{O}\left( k_F^2 \right)
 \right\}.
 \end{equation}
 We note that the force strength $V_\mathrm{GMB}$ depends on the Fermi momentum $k_F$. Expanded
 in powers of $k_F$, relevant to the low-density limit $k_F \rightarrow 0$ is the
 linear term in $k_F$. 
 It can be expressed also in terms of  the density $\rho=k_F^3/3\pi^2$ as
 \begin{equation}
 V_\mathrm{GMB}=V_0\left\{ 1 - \eta \left(\frac{\rho}{\rho_0}\right)^{1/3} 
 + \mathrm{O}\left( \left(\frac{\rho}{\rho_0}\right)^{2/3} \right)
 \right\}
 \end{equation}
 with
\begin{equation}
  \eta=\frac{1+2\log 2}{3}\frac{k_{F0}}{k_\mathrm{cut} - \frac{\pi}{2a}}, \ \ \ \ k_{F0}=(3\pi^2\rho_0)^{1/3}.
\end{equation}

\section*{Appendix B:  BCS calculation for uniform neutron matter}

Here we describe the selfconsistent Hartree-Fock plus BCS approximation
which is adopted to describe the pairing property of uniform neutron matter. 

For a given value of the neutron Fermi  energy $\lambda_n$, we numerically solve the coupled equations
\begin{equation}
 \Delta(\lambda_n)=-\frac{V_n[\rho_n]}{4\pi^2} \int_0^{k'_c} dk k^2 \frac{\Delta(\lambda_n)}{E(k)},
\end{equation}
\begin{equation}
E(k)= \sqrt{(e(k)-\lambda_n)^2+ \Delta^2}, \ \ \ e(k) = \frac{\hbar^2 k^2}{2m_n^*(\rho_n)} + U_n(\rho_n),
\end{equation}
\begin{equation}
\rho_n(\lambda_n) =\frac{1}{2\pi^2} \int_0^{k'_c} dk k^2 \left\{ 1+ \frac{e(k)-\lambda_n}{E(k)} \right\},
\label{rhon}
\end{equation}
where $U_n(\rho_n)$ and $m_n^*(\rho_n)$ are the Hartree-Fock potential and the effective mass
of neutrons, obtained from the SLy4 functional.\footnote{In our previous publication \cite{inakura2017}, we made an approximation to Eq.  (\ref{rhon}) 
using a Fermi gas relation $\rho_n =\frac{1}{3\pi^2}\left(\frac{2m\lambda_n}{\hbar^2}\right)^{3/2}$. Hence the result for neutron matter 
shown in Fig. 2 of Ref. \cite{inakura2017} is slightly different from that of the present study.}
The cut-off momentum $k'_c$ is determined by $e(k'_c)-\lambda=E_\mathrm{cut}$ so that it
corresponds to the cut-off energy in the coordinate-space HFB calculation. 
The above scheme is called the uniform-BCS calculation in this paper.

The coherence length $\xi$ can be calculated by evaluating the size of the Cooper pair 
and is given
  \begin{align}
  \xi=\sqrt{\left<r^2\right>}
 \end{align}
 \begin{align}
  \left<r^2\right>=\int d \vec{r} r^2\left|\Psi_\mathrm{pair}(r)\right|^2=
\frac{\int^\infty_0 dk k^2 (\frac{\partial}{\partial k} u_k v_k)^2}{\int^\infty_0 dk k^2 ( u_k v_k)^2}.
 \end{align}

\let\doi\relax

 \end{document}